\documentstyle[multicol,aps,epsf]{revtex}

\begin{document}
\title{Fate of Zero-Temperature Ising Ferromagnets}
\author{V.~Spirin, P.~L.~Krapivsky, and S.~Redner}

\address{Center for BioDynamics, Center for Polymer Studies, 
and Department of Physics, Boston University, Boston, MA, 02215}
\maketitle
\begin{abstract}
  
  We investigate the relaxation of homogeneous Ising ferromagnets on finite
  lattices with zero-temperature spin-flip dynamics.  On the square lattice,
  a frozen two-stripe state is apparently reached approximately 1/4 of the
  time, while the ground state is reached otherwise.  The asymptotic
  relaxation is characterized by two distinct time scales, with the longer
  stemming from the influence of a long-lived diagonal stripe ``defect''.  In
  greater than two dimensions, the probability to reach the ground state
  rapidly vanishes as the size increases and the system typically ends up
  wandering forever within an iso-energy set of stochastically ``blinking''
  metastable states.

\end{abstract}  
\begin{multicols}{2}
  
  What happens when an Ising ferromagnet, with spins endowed with Glauber
  dynamics\cite{glauber}, is suddenly cooled from high temperature to zero
  temperature ($T=0$)?  A first expectation is that the system should
  coarsen\cite{rev} and eventually reach the ground state.  However, even the
  simple Ising ferromagnet admits a large number of metastable states with
  respect to Glauber spin-flip dynamics.  Therefore at zero temperature the
  system could get stuck forever in one of these states.
  
  In this Letter, we present evidence that the behavior of such a kinetic
  Ising model is richer than either of these scenarios.  While the ground
  state is always reached in one dimension, there appears to be a non-zero
  probability that the square lattice system freezes into a ``stripe'' phase,
  at least for equal initial concentrations of $\uparrow$ and $\downarrow$
  spins\cite{other}.  The relaxation is governed by two distinct time scales,
  the larger of which stems from a long-lived diagonal stripe ``defect''.  On
  hypercubic lattices ($d\geq 3$), the probability to reach the ground state
  vanishes in the thermodynamic limit and the system ends up wandering
  forever on an iso-energy subset of connected metastable states.  Again, the
  relaxation seems to be characterized by at least two time scales.  It bears
  emphasizing that these long-time anomalies require that the limit $T\to 0$
  is taken {\em before} the thermodynamic limit $L\to\infty$; very different
  behavior occurs if $L\to\infty$ before $T\to 0$\cite{ns}.
    
  We can easily appreciate the peculiarities of zero-temperature dynamics for
  odd-coordinated lattices, such as the honeycomb lattice.  Here any
  connected cluster in which each spin has at least 2 aligned neighbors is
  energetically stable in a sea of opposite spins.  For {\em any\/} initial
  state, a sufficiently large system will have many such stable defects.
  Because the number of such metastable states generally scales exponentially
  with the total number of spins $N$, the system necessarily freezes into one
  of these states.  However, on even-coordinated lattices the number of
  metastable states grows as a slower, stretched exponential function of $N$,
  and they affect the asymptotic relaxation in much more subtle way.
  
  We therefore study the homogeneous Ising model, with Hamiltonian ${\cal
    H}=-J\sum_{\langle ij\rangle}\sigma_i\sigma_j$, where $\sigma_i=\pm 1$
  and the sum is over all nearest-neighbor pairs of sites $\langle i
  j\rangle$.  We assume initially uncorrelated spins, with $\sigma_j(t=0)=\pm
  1$ equiprobably, which evolve by zero-temperature Glauber
  dynamics\cite{glauber}, corresponding to a quench from $T=\infty$ to $T=0$.
  We focus on $d$-dimensional hypercubic lattices with linear size $L$ and
  periodic boundary conditions.  Most of our results continue to hold for
  free boundary conditions and on arbitrary even-coordinated lattices.
  
  Glauber dynamics at zero temperature involves picking a spin at random and
  considering the energy change $\Delta E$ if the spin were flipped.  If
  $\Delta E<0$ ($>$0), the flip is accepted (rejected), while if $\Delta
  E=0$, the attempt is accepted with probability 1/2.  After each event, time
  is updated by $1/L^d$, so that each spin undergoes, on average, one update
  attempt in a single time unit.  In practice, we update only the flippable
  spins (those with $\Delta E\leq 0$) and update the time by $1/({\rm number\ 
    of\ flippable\ spins})$.  For each initial state, one realization of the
  dynamics is run until the final state.  At $T=0$, metastable states in this
  dynamics have an infinite lifetime and these can prevent the equilibrium
  ground state from being reached.  This is the basic reason why dynamics at
  $T=0$ is very different from that of small positive temperature.
  
  In one dimension, it is easy to determine the ultimate fate of the
  system\cite{liggett}.  In $T=0$ Glauber kinetics, the expectation value of
  the $i^{\rm th}$ spin, $s_i\equiv \langle\sigma_i\rangle$, obeys the
  diffusion equation and therefore the average magnetization $\langle
  m\rangle={1\over L} \sum_j s_j$ is conserved\cite{glauber}.  Since there
  are no metastable states in one dimension, the only possible final states
  are all spins up or all spins down.  For initial magnetization $m(0)$, a
  final magnetization $m(\infty)=m(0)$ can be achieved only if a fraction
  ${1\over 2}(1+m(0))$ of all realizations of the dynamics ends with all
  spins up and a fraction ${1\over 2}(1-m(0))$ with all spins down.
  
  On the square lattice, there exists a huge number of metastable states
  which consist of alternating vertical (or horizontal) stripes whose widths
  are all $\geq 2$.  These arise because in zero-temperature Glauber dynamics
  a straight boundary between up and down phases is stable; a reversal of any
  spin along the boundary increases its length and raises the energy.  Note
  that a stripe of width one is not stable because it can be cut in two at no
  energy cost by flipping one of the spins in the stripe.
  
\begin{figure}
  \narrowtext \epsfxsize=2.2in\hskip 0.4in\epsfbox{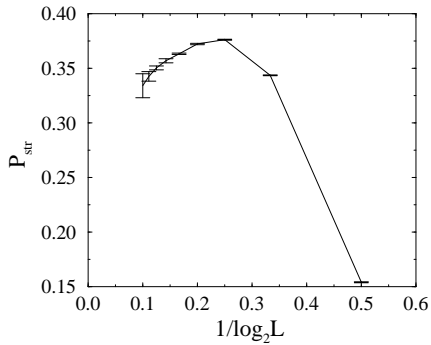} \vskip
  0.15in
\caption{Probability that an $L\times L$ square eventually reaches a stripe
  state as a function of $1/\log_2 L$ for $L$ up to 1024.  Each data point
  (with error bars) is based on $10^5$ initial spin configurations for $L\leq
  256$, and then 13,500 and 2,700 configurations respectively for $L=512$ and
  $L=1024$. 
\label{stripe}}
\end{figure}

The mere existence of these metastable states implies that a finite sample
may not reach the ground state.  However, one might expect that the
probability to reach the ground state approaches unity as the system size
grows: $\lim_{L\to\infty}P_{\rm gs}=1$.  Our numerical simulations on
$L\times L$ squares appear to disagree with this expectation
(Fig.~\ref{stripe}).  The probability $P_{\rm gs}$ grows extremely slowly
with $L$ and extrapolates to a value of approximately $3/4$ as $L\to\infty$,
so that the probability to reach a two-stripe state, $P_{\rm str}=1-P_{\rm
  gs}$ would be non-zero.  We also find that states with more than two
stripes almost never appear.

\begin{figure}
  \narrowtext \hskip -0.1in\epsfxsize=1.67in\epsfysize=1.74in\hskip 0.0in\epsfbox{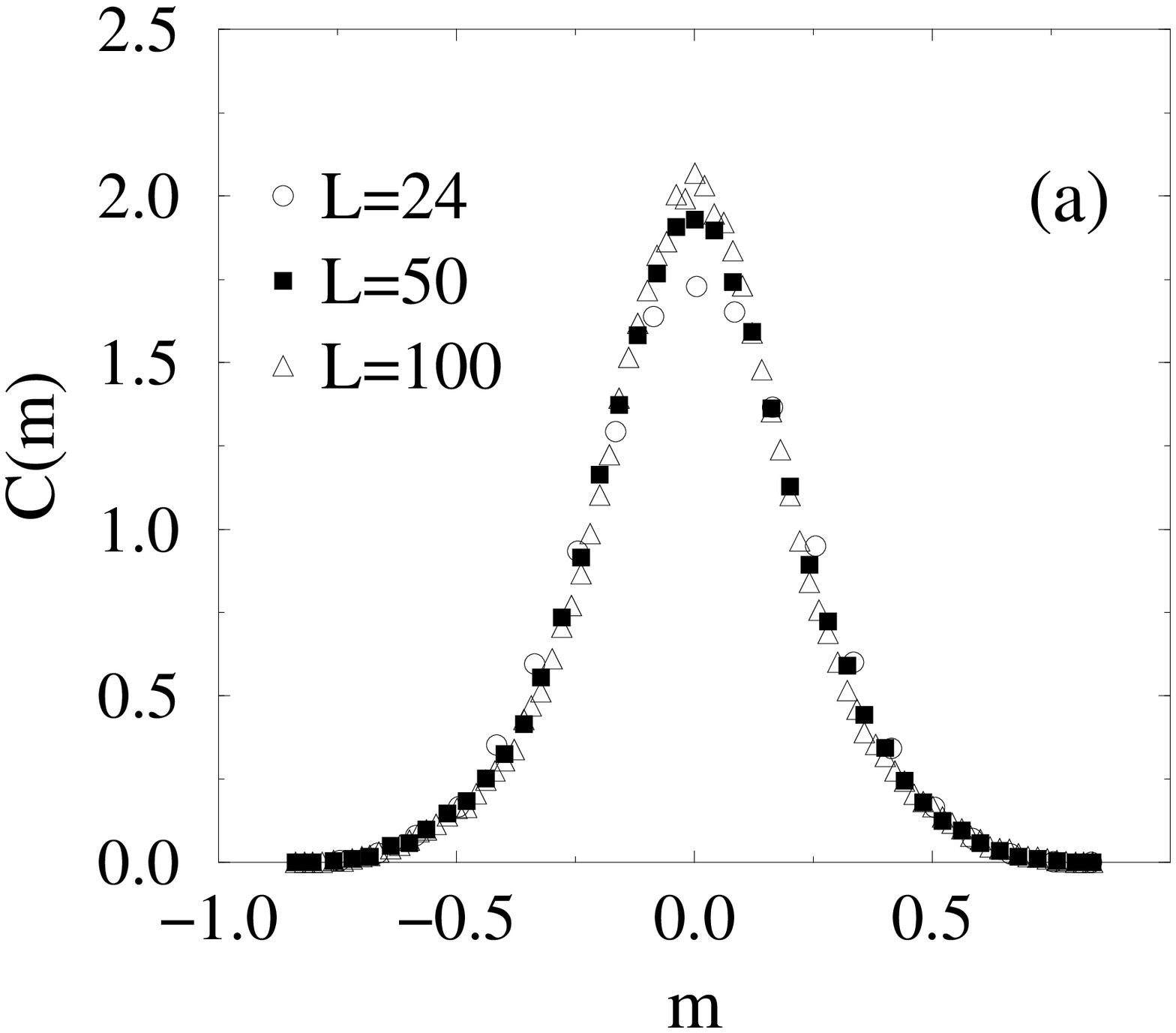} 
  \hskip 0.03in\epsfxsize=1.55in\epsfysize=1.80in\hskip 0.0in\epsfbox{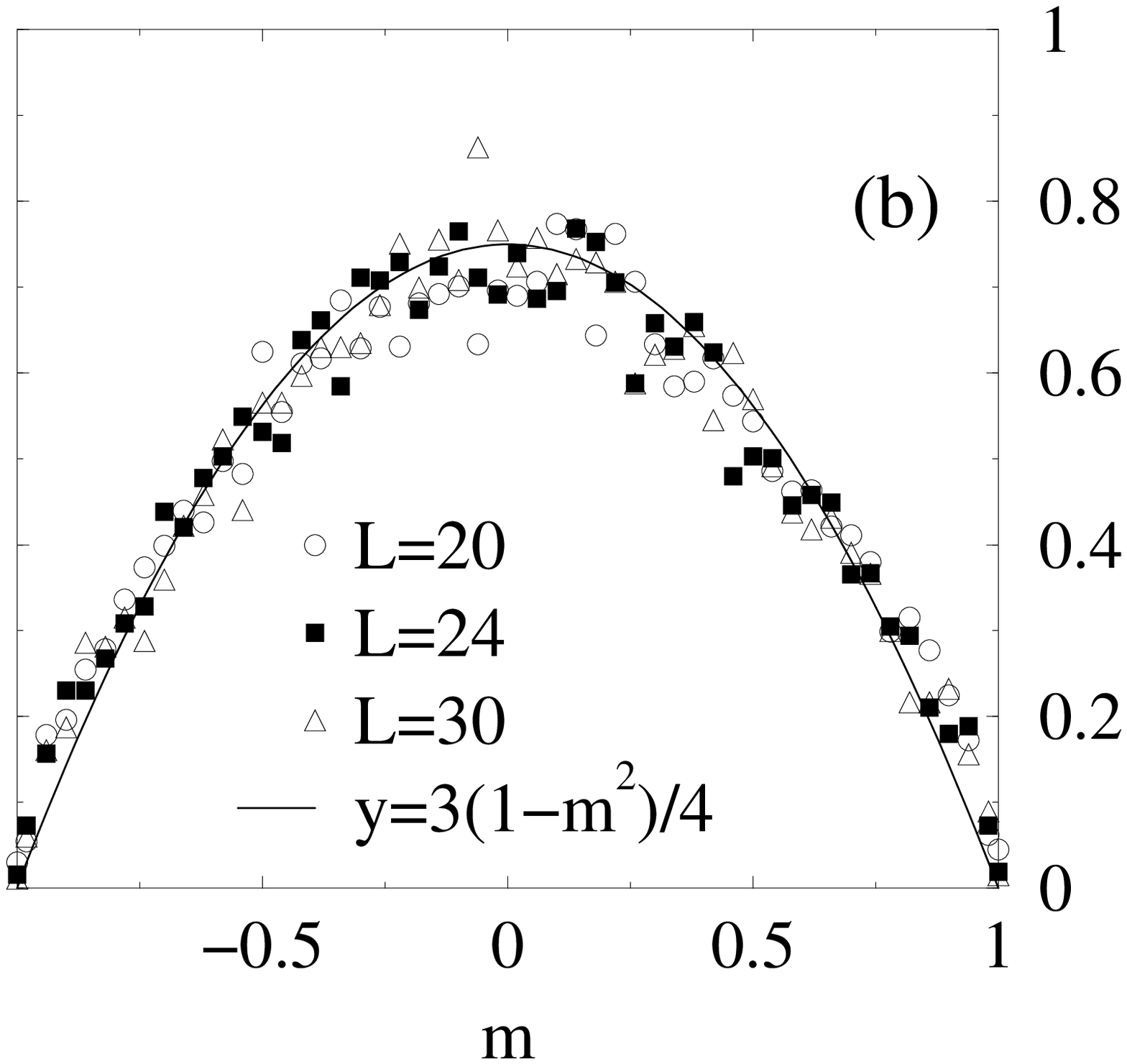} 
\vskip
  0.15in
\caption{Final magnetization distribution on the 
  (a) square, and (b) cubic lattices.  On the square lattice, this
  distribution narrows as $L$ increases, but appears to reach a non-singular
  limit.  On the cubic lattice, data collapse occurs even at small sizes.
  The number of realizations is $10^5$ for the square and $\geq 10^4$ for the
  cubic lattices.
\label{m}}
\end{figure}

When the two-stripe state is reached on the square lattice, both stripes have
width typically of the order of $L/2$, as seen by a gradual narrowing of the
continuous component $C_{\rm 2d}(m)$ of the final magnetization distribution,
$F_{2d}(m)= {1\over 2} P_{\rm gs} \left[\delta(m-1)+\delta(m+1)\right]+P_{\rm
  str}C_{2d}(m)$ (Fig.~\ref{m}(a)).  However, as $L$ increases the
magnetization distribution appears to converge to a finite-width scaling
limit.  On the simple cubic lattice, there are many more metastable state
topologies and also relatively more states with narrow stripes, so that there
is a larger probability that the final magnetization is close to $\pm 1$.
The final magnetization distribution also exhibits good data collapse even at
relatively small system sizes.  Strikingly, the final magnetization
distribution on the cubic lattice is well fit by $F_{3d}(m)={3\over
  4}(1-m^2)$ (Fig.~\ref{m}(b)).

\begin{figure}
  \narrowtext \epsfxsize=2.9in \hskip 0.05in\epsfbox{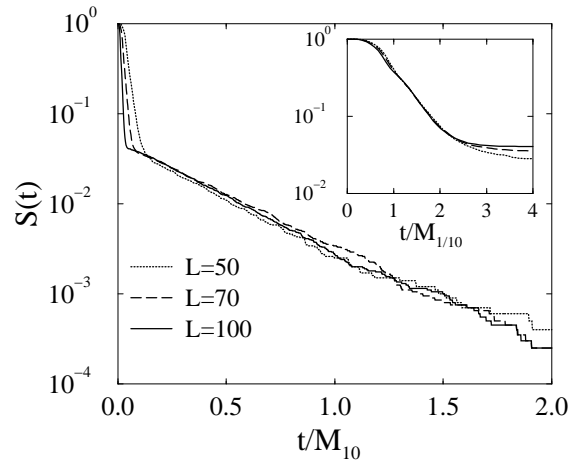} \vskip
  0.15in
\caption{Time dependence of the survival probability $S(t)$ on $L\times L$ 
  squares.  Main graph: $S(t)$ versus $t/M_{10}$ to highlight the long-time
  exponential tail, where $M_k\equiv\langle t^k\rangle^{1/k}$ is the $k^{\rm
    th}$ reduced moment of the time to reach the final state.  Scaling sets
  in after $S(t)$ has decayed to approximately 0.04.  Inset: $S(t)$ versus
  $t/M_{1/10}$ to highlight the scaling and the faster exponential decay in
  the intermediate-time regime.
\label{surv}}
\end{figure}

Intriguing behavior is also exhibited by the survival probability $S(t)$ that
the system has not yet reached its final state by time $t$.  As shown in
Fig.~\ref{surv}, $S(t)$ is controlled by two different time scales.  On a
semi-logarithmic plot, $S(t)$ lies on one straight line with large negative
slope for intermediate times and crosses over to another line with smaller
negative slope at long times.  In this intermediate time regime, the energy
decays as $t^{-1/2}$, as expected\cite{rev} The crossover in $S(t)$ occurs
when domains, which grow according to the classical $t^{1/2}$ law\cite{rev},
reach the system size.  This gives the crossover time $\tau_c\propto L^2$.

\begin{figure}
  \narrowtext \epsfxsize=2.2in \hskip 0.4in\epsfbox{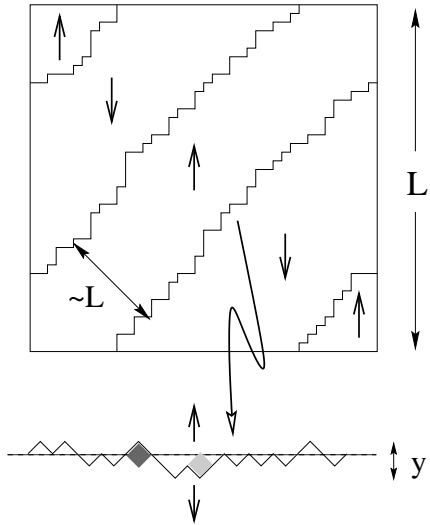} \vskip
  0.15in
\caption{Diagonal stripe configuration on the square lattice.
  The lower portion shows part of one interface rotated by $45^\circ$.
  Zero-temperature Glauber dynamics allows for either particle deposition at
  the bottom of a valley (light-shaded square) -- corresponding to the
  spin-flip even $\uparrow\to\downarrow$ -- or particle evaporation from a
  peak (filled square) -- corresponding to $\downarrow\to\uparrow$.
\label{diagonal}}
\end{figure}

Quite surprisingly, the source of the long-time anomaly in $S(t)$ arises from
the approximately 4\% of the configurations in which a diagonal stripe
appears (Fig.~\ref{diagonal}).  On the torus, this configuration consists of
one stripe of $\uparrow$ spins and another of $\downarrow$ spins which, by
symmetry, have width of order $L$.  Each of these stripes winds once in both
toroidally and poloidally on the torus; they cannot evolve into straight
stripes by a continuous deformation of the boundaries.  Consequently a
diagonal stripe configuration should ultimately reach the ground state.

Diagonal stripes are also extremely long-lived.  For $L=200$, for example,
the time for such a configuration to reach the final state is two orders of
magnitude larger than the typical time.  To understand this long lifetime, we
view a diagonal boundary as an evolving interface in a reference frame
rotated by $45^\circ$\cite{prl}.  

In this frame (Fig.~\ref{diagonal} lower), a spin flip is equivalent to a
``particle'' which either deposits at the bottom of a valley
($\uparrow\to\downarrow$) or evaporates from a peak
($\downarrow\to\uparrow$).  In a single time step each such event occurs with
probability 1/2.  For an interface with transverse dimension of order $L$,
let us assume that there are of the order of $L^\mu$ such height extrema.
Ref.~\cite{prl} predicts $\mu=1$, but we temporarily keep the value arbitrary
for clarity.  Accordingly, in a single time step, where all interface update
attempts occur once on average, the interface center-of-mass moves a distance
$\Delta y\sim L^{\mu/2}/L$.  This gives an interface diffusivity $D\sim
(\Delta y)^2\sim L^{\mu-2}$.  We then estimate the lifetime $\tau_{\rm diag}$
of a diagonal stripe as the time for the interface to move a distance of
order $L$ to meet another interface.  This gives $\tau_{\rm diag}\sim
L^2/D\sim L^{4-\mu}$.  Using the results of Ref.~\cite{prl}, we expect
$\tau_{\rm diag}\propto L^3$.

\begin{figure}
  \narrowtext \epsfxsize=2.4in\hskip
  0.3in\epsfbox{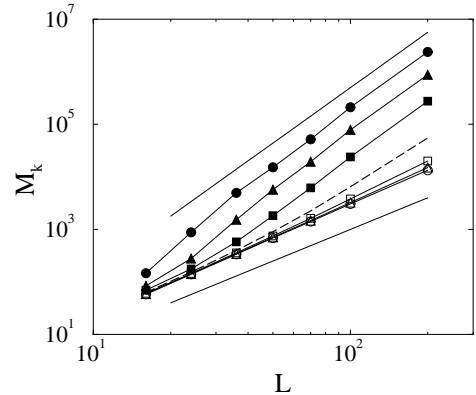} \vskip 0.15in
\caption{Dependence of  $M_k\equiv \langle t^k\rangle^{1/k}$ on $L$.  
  Shown are the cases $k=1$ (dashed curve), $k=1/2$ ($\Box$), 1/4 ($\Delta$),
  and 1/10 ($\circ$), as well as $k=2$, 4, and 10 (corresponding filled
  symbols).  The thin straight lines have slopes 2 and 3.5.
\label{moments}}
\end{figure}

The survival probability reflects these two time scales (Fig.~\ref{surv}),
and their $L$ dependence is cleanly visible in the reduced moments $M_k\equiv
\langle t^k\rangle^{1/k}$ of the time until the final state is reached.  The
main contribution to the moments with $k<1$ comes from short-lived
configurations, while for $k>1$ the main contribution comes from long-lived
diagonal-stripe configurations.  Our data for $M_k$ with $k<1$ scales
approximately as $L^2$, while for $k>1$, $M_k$ scales roughly as $L^{3.5}$,
somewhat faster growth than expected (Fig.~\ref{moments}).

In greater than two dimensions, the probability to reach the ground state
rapidly vanishes as the system size increases.  For example, $P_{\rm
  gs}\approx 0.04$ and $0.003$ for cubic lattices of linear dimension $L=10$
and 20.  For larger lattices, the ground state has not been reached in any of
our simulations.  One obvious reason why the system ``misses'' the ground
state is the rapid increase in the number of metastable states with spatial
dimension.  This proliferation of metastable states makes it more likely that
a typical configuration will eventually reach one of these states rather than
the ground state.  Another striking feature is that many metastable states in
three dimensions form connected iso-energy sets, while metastable states are
all isolated in two dimensions.  Thus a three-dimensional system can end up
wandering forever on one of these connected sets.

\begin{figure}
  \narrowtext \epsfxsize=2.in \hskip 0.6in\epsfbox{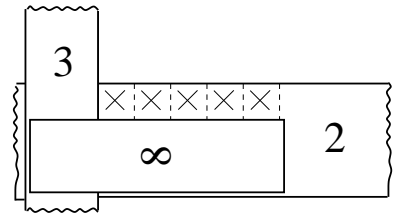} \vskip
  0.15in
\caption{A stochastic blinker on the cubic lattice.
\label{blinkers}}
\end{figure}

A specific example from a simulation on a small cube is sketched in
Fig.~\ref{blinkers}.  By viewing the spins as cubic blocks, the cluster of
aligned spins appears as a ``building'' with a $2$-storey section (marked 2),
an adjacent $3$-storey section, and a section (marked $\infty$) which wraps
around the torus in the vertical direction and rejoins the building on the
ground floor.  The wiggly lines indicate that building sections also wrap
around in the $x$- and $y$-directions.

The sites marked by $\times$ are ``blinkers''.  Consider the leftmost such
site.  Since there are three directions where the nearest neighbors are part
of the building, this leftmost spin can flip with zero energy cost.  If this
occurs, then its right neighbor, which was initially stable, can now flip
with no energy cost.  This sequence can continue until the right edge of the
$\infty$ section of the building but no further.  Therefore this third-storey
addition performs a random walk, constrained to move forever in the interval
marked by the $\times$ sites.  While this construction appears idiosyncratic,
such a structure is the basic element of generic blinkers on even-coordinated
Cayley trees.

Another feature of the final state is that it almost always consists of only
two interpenetrating clusters which both percolate in all three Cartesian
directions.  These two percolating clusters must each contain no convex
corners to be stable at $T=0$.  While there are also metastable states with
many components and metastable states with components percolating in one or
in two directions, such configurations are generally not reached when the
system is large enough.

The energy decay on the cubic lattice also suggests that there exists more
than one relaxational time scale.  Initially, the energy decreases
systematically in a manner consistent with a power-law decay.  At longer
times, however, the energy exhibits plateaux of increasing duration,
punctuated by small energy decreases.  Ultimately the final energy is reached,
after which constant-energy stochastic blinking occurs {\it ad infinitum}.
Our data for the time until the appearance of the first energy plateau scales
roughly as $L^3$, while the time to reach the final energy seems to increase
faster than any power of $L$.

The phase-space structure of the metastable states appears to be a crucial
element in understanding the fate of Ising ferromagnets.  The simplest aspect
is to estimate the number of metastable states $M_d(N)$ as a function of the
spatial dimension $d$ and number of spins $N=L^d$.  In two dimensions, a
metastable state contains alternating horizontal or vertical stripes of up
and down spins, with each stripe of width $\geq 2$.  This is identical to the
number of ground states of a periodic Ising chain with nearest-neighbor
ferromagnetic interaction $J_1$ and second-neighbor antiferromagnetic
interaction $J_2$ (axial next-nearest neighbor Ising (ANNNI) model), when
$J_2=-J_1/2$.  For the chain with open boundaries, the number of metastable
state was previously found in terms of the Fibonacci numbers\cite{red}.  For
the periodic system the number of metastable states is
\begin{equation}
\label{eq:M}
M_2={g^{L-1}-(-g)^{-L+1}\over \sqrt{5}}-{2\over \sqrt{3}}\,
\sin{\pi\over 3}(L-1)+2,
\end{equation}
where $g={1\over 2}(1+\sqrt{5})$ is the golden ratio, and the first term is
just the ANNNI model degeneracy.  Eq.~(\ref{eq:M}) therefore gives
$M_2(V)\sim e^{A_2\sqrt{N}}$ with $A_2=\ln g$.

For $d=3$, we give a lower bound for the number of metastable states by
generalizing the stripe states of the square lattice.  Consider states which
consist of an array of straight filaments such that each filament
cross-section is rectangular and the ``Manhattan'' distance between any two
rectangles is $\geq 2$.  The number of packings of such filaments is of the
order of $\exp(cL^2)$, where $c$ is a constant.  This gives the lower bound
for the number of metastable states, $M_3(N)>\exp(cN^{2/3})$.  This same
construction in $d$ dimensions gives $M_d(N)>\exp(cN^{(d-1)/d})$.  While we
have not succeeded in constructing an upper bound, it seems plausible that
this bound has the same functional form as the lower bound; hence $M_d(N)\sim
\exp\left(A_d N^{(d-1)/d}\right)$.  In addition, it can be verified that the
number of metastable states on the Cayley tree grows exponentially with the
number of spins.  Thus metastable states become relatively more numerous as
the dimension increases and their influence on long-time kinetics should
correspondingly increase.

In summary, the homogeneous Ising ferromagnet exhibits surprisingly rich
behavior following a quench from infinite to zero temperature.  On the square
lattice, there appears to be a non-zero probability of reaching a static
two-stripe state.  Evolution via a diagonal stripe configuration is
responsible for a two-time-scale relaxation kinetics.  On the cubic lattice,
the probability of reaching both the ground state or a frozen metastable
state vanishes rapidly as the system size increases.  The system instead
reaches a finite iso-energy attractor of metastable states upon which it
wanders stochastically forever.

\medskip 

We thank C. Godr\`eche, M. Grant, J. Krug, T. Liggett, C. Newman, Z. R\'acz,
J. Sethna, and D. Stein for helpful discussions, and NSF grant DMR9978902 for
financial support.

\end{multicols} 
\end{document}